\documentclass[aps,showpacs,amsfonts,nofootinbib,preprintnumbers,nobalancelastpage]{revtex4}
\usepackage{graphicx}
\usepackage{epsfig}
\usepackage{xcolor}
\usepackage{rotating}
\usepackage{amsmath}
\usepackage{braket}
\usepackage{caption}       
\usepackage{multirow}      
\usepackage{booktabs}      
\usepackage{array}         

\newcommand{\Frac}[2]{\frac{\displaystyle #1}{\displaystyle #2}}

\newcommand{\UH}{\mathbf{U}}\newcommand{\DLR}{\mathbf{D}}
\def\Tr{\hbox{Tr}}

\begin{document}


\title{Unitarity Constraints on Anomalous Quartic Couplings}

\author{Eduardo da Silva Almeida}
\email{eduardo.silva.almeida@usp.br}
\affiliation{Instituto de F\'{\i}sica,
             Universidade de S\~ao Paulo, S\~ao Paulo -- SP, Brazil.}

\author{O.\ J.\ P.\ \'Eboli}
\email{eboli@if.usp.br}
\affiliation{Instituto de F\'{\i}sica,
             Universidade de S\~ao Paulo, S\~ao Paulo -- SP, Brazil.}

\author{M.\ C.\ Gonzalez--Garcia} \email{concha@insti.physics.sunysb.edu}
\affiliation{%
  Instituci\'o Catalana de Recerca i Estudis Avan\c{c}ats (ICREA),}
\affiliation {Departament d'Estructura i Constituents de la Mat\`eria, 
Universitat
  de Barcelona, 647 Diagonal, E-08028 Barcelona, Spain}
\affiliation{%
  C.N.~Yang Institute for Theoretical Physics, SUNY at Stony Brook,
  Stony Brook, NY 11794-3840, USA}

\begin{abstract}
  We obtain the partial-wave unitarity constraints on the
  lowest-dimension effective operators which generate anomalous
  quartic gauge couplings but leave the triple gauge couplings
  unaffected.  We consider operator expansions with linear and
  nonlinear realizations of the electroweak symmetry and explore the
  multidimensional parameter space of the coefficients of the relevant
  operators: 20 dimension-eight operators in the linear expansion and
  5 ${\cal O}(p^4)$ operators in the derivative expansion.  We study
  two-to-two scattering of electroweak gauge bosons and Higgs bosons
  taking into account all coupled channels and all possible helicity
  amplitudes for the $J=0,1$ partial waves.  In general, the bounds
  degrade by factors of a {\sl few} when several operator coefficients
  are considered to be nonvanishing simultaneously. However, this
  requires considering constraints from both $J=0$ and $J=1$ partial
  waves for some sets of operators.
\end{abstract}

\preprint{YITP-SB-2020-8}

 \maketitle
\renewcommand{\baselinestretch}{1.15}
%

\section{Introduction}

The structure of the triple (TGC) and quartic (QGC) electroweak
gauge-boson interactions in the Standard Model (SM) is determined by
the gauge symmetry $SU(2)_L \otimes U(1)_Y$.  Therefore, it is
important to measure both TGC and QGC, not only to further test the SM
or have indications of new physics, but also to determine whether the
gauge symmetry is realized linearly or nonlinearly in the low-energy
effective theory of the electroweak symmetry breaking
sector~\cite{Brivio:2013pma}. \smallskip

Generically, deviations from the SM predictions for TGC and QGC are
generated by higher-order operators parametrizing indirect effects of
new physics.  Collider experiments probe TGC in the pair production of
electroweak gauge bosons while the study of QGC requires the
production of three electroweak vector bosons,  the exclusive
production of gauge-boson pairs or the vector-boson-scattering
production of electroweak vector boson pairs~\cite{Belanger:1992qh,
  Belanger:1992qi, Eboli:2000ad,Eboli:2003nq, Chatrchyan:2013akv,
  Chatrchyan:2014bza, Aad:2015uqa, Khachatryan:2016mud,
  Khachatryan:2016vif, Aaboud:2016ffv, Khachatryan:2017jub,
  Aaboud:2017tcq, Sirunyan:2017lvq, Sirunyan:2017fvv,
  Sirunyan:2019der, Sirunyan:2020tlu}.  Therefore, the Wilson
coefficients of effective operators that contain both TGC and QGC are
more strongly constrained through the study of their TGC component.
\smallskip

For this reason most of present LHC searches for effects of QGC focus
on the so-called {\sl genuine} QGC operators, that is, operators
generating QGC but that do not have any TGC associated with them.  In
a scenario where the $SU(2)_L \otimes U(1)_Y$ is realized linearly the
lowest-order QGC are given by dimension-eight
operators~\cite{Eboli:2006wa}. Alternatively, if the gauge symmetry is
implemented nonlinearly the lowest-order QGC appear at
${\cal O}(p^4)$~\cite{Alonso:2012px,Brivio:2013pma}.  \smallskip

It is well known that departures of the TGC and QGC from the SM
predictions lead to the growth of scattering
amplitudes~\cite{Cornwall:1974km}, signalizing the existence of new
physics. Thus, when probing anomalous QGC one must verify whether
perturbative partial-wave unitarity is satisfied to guarantee
consistency of the analyses.  This is all well established and it has
been previously addressed in the literature~\cite{Boos:1997gw,
  Eboli:2000ad, Guo:2019agy, Guo:2020lim, Arnold:2008rz,
  Kalinowski:2018oxd, Kozow:2019txg, Chaudhary:2019aim}.  It is also
implemented in some form in the QGC searches by both ATLAS and CMS
collaborations, see for instance Refs.
~\cite{Khachatryan:2016mud,Khachatryan:2016vif,
  Aaboud:2016ffv,Khachatryan:2017jub, Aaboud:2017tcq,Sirunyan:2017fvv,
  Sirunyan:2020tlu}, either by introducing {\em ad-hoc} form factors
or unitarization procedures (see Ref.~\cite{Garcia-Garcia:2019oig} for
a study of the dependence on the unitarization procedure employed), or
by directly evaluating the maximum center-of-mass energy allowed by
unitarity as obtained from the VBFNLO framework~\cite{Arnold:2008rz}.
However, these unitarity studies are not complete since they consider
just a few scattering channels, or a limited set of QGC effective
operators, or they restricted the analysis to the $J=0$ partial
wave. \smallskip

In this work we complement the existing literature on the subject by
systematically presenting the unitarity bounds in the multidimensional
parameter space of the coefficients of the relevant operators in both
linear and nonlinear realizations of the electroweak symmetry.  We
study two-to-two scattering of electroweak gauge bosons and Higgs
bosons taking into account all coupled channels and all possible
helicity amplitudes for the $J=0,1$ partial waves. Indeed we find that
$J=1$ partial-wave unitarity effects are relevant to derive the most
stringent limits in some scenarios when the effects of several
operators are considered simultaneously, \smallskip


This paper is organized as follows: we present in
Section~\ref{sec:forma} the QCG operators that we consider in our
analyses, as well as basic expressions of partial-wave unitarity
needed for our studies. Section~\ref{sec:results} contains our results
that are discussed in Section~\ref{sec:discussion}. 
\smallskip

\section{Analyses framework}
\label{sec:forma}

Here, we introduce the effective interactions considered in this work,
as well as the unitarity relations that we use to constrain
them. \smallskip

\subsection{Effective Lagrangian}
\label{sec:lag}

\subsubsection{Linear realization of the gauge symmetry}

Assuming that the new state observed in 2012 is in fact the SM Higgs
boson and that it belongs to an electroweak scalar doublet, we can
construct a low-energy effective theory where the
$SU(2)_L \otimes U(1)_Y$ gauge symmetry is linearly
realized~\cite{Buchmuller:1985jz, Leung:1984ni, DeRujula:1991ufe,
  Hagiwara:1993ck, GonzalezGarcia:1999fq, Grzadkowski:2010es,
  Passarino:2012cb} which takes the form
\begin{equation}
{\cal L}_{\rm eff} = {\cal L}_{\text SM} + \sum_{n=5}^\infty\sum_i
\frac{f^{(n)}_i}{\Lambda^{n-4}} {\cal O}^{(n)}_i \;\; ,
\label{l:eff}
\end{equation}
where the dimension--n operators ${\cal O}^{(n)}_i$ involve
gauge bosons, Higgs doublets, fermionic fields, and
covariant derivatives of these fields. Each operator has a
corresponding Wilson coefficient $f^{(n)}_i$ and $\Lambda$ is the
characteristic energy scale at which new physics (NP) becomes
apparent. \smallskip

Here,  we are interested in operators that lead to QGC without a TGC
counterpart.  The lowest dimension of such genuine QGC operators is
eight~\cite{Eboli:2006wa}.  In what follows, we consider the bosonic
dimension-eight operators relevant to two-to-two scattering processes
involving Higgs and/or gauge bosons at tree level, and that conserve
$C$ and $P$~\cite{Eboli:2016kko}. Moreover, we classify them by the
number of gauge-boson strength fields contained in the
operator. \smallskip

In the first class of genuine QGC, the operators contain just covariant
derivatives of the Higgs field:
\begin{equation}
\begin{array}{lll}
  {\cal O}_{S,0} = 
\left [ \left ( D_\mu \Phi \right)^\dagger
 D_\nu \Phi \right ] \times 
\left [ \left ( D^\mu \Phi \right)^\dagger
D^\nu \Phi \right ]
&\hbox{   ,   }
&
{\cal O}_{S,1} =
 \left [ \left ( D_\mu \Phi \right)^\dagger
 D^\mu \Phi  \right ] \times
\left [ \left ( D_\nu \Phi \right)^\dagger
D^\nu \Phi \right ]
\; , 
\\
&&
\\
  {\cal O}_{S,2} =
 \left [ \left ( D_\mu \Phi \right)^\dagger
 D_\nu \Phi  \right ] \times
\left [ \left ( D^\nu \Phi \right)^\dagger
D^\mu \Phi \right ]
\; ,
\end{array}
\label{eq:dphi}
\end{equation}
where $\Phi$ stands for the Higgs doublet, the covariant derivative is
given by
$D_\mu \Phi = (\partial_\mu + i g W^j_\mu \frac{\sigma^j}{2} + i
g^\prime B_\mu \frac{1}{2}) \Phi$ and $\sigma^j$ ($j=1,2,3$) represent
the Pauli matrices. \smallskip

In the second class of genuine QGC the operators exhibit two covariant
derivatives of the Higgs field, as well as two field strengths:
\begin{equation}
\begin{array}{lcll}
 {\cal O}_{M,0} =   \hbox{Tr}\left [ \widehat{W}_{\mu\nu} \widehat{W}^{\mu\nu} \right ]
\times  \left [ \left ( D_\beta \Phi \right)^\dagger
D^\beta \Phi \right ]
&,& 
 {\cal O}_{M,1} 
=   \hbox{Tr}\left [ \widehat{W}_{\mu\nu} \widehat{W}^{\nu\beta} \right ]
\times  \left [ \left ( D_\beta \Phi \right)^\dagger
D^\mu \Phi \right ]
&,
\\
 {\cal O}_{M,2} =   \left [ B_{\mu\nu} B^{\mu\nu} \right ]
\times  \left [ \left ( D_\beta \Phi \right)^\dagger
D^\beta \Phi \right ]
&,&
 {\cal O}_{M,3} =   \left [ B_{\mu\nu} B^{\nu\beta} \right ]
\times  \left [ \left ( D_\beta \Phi \right)^\dagger
D^\mu \Phi \right ]
&,
\\
  {\cal O}_{M,4} = \left [ \left ( D_\mu \Phi \right)^\dagger \widehat{W}_{\beta\nu}
 D^\mu \Phi  \right ] \times B^{\beta\nu}
&,&
  {\cal O}_{M,5} = \left [ \left ( D_\mu \Phi \right)^\dagger \widehat{W}_{\beta\nu}
 D^\nu \Phi  \right ] \times B^{\beta\mu}+ {\rm h.c.}
&,
\\
  {\cal O}_{M,7} = \left [ \left ( D_\mu \Phi \right)^\dagger \widehat{W}_{\beta\nu}
\widehat{W}^{\beta\mu} D^\nu \Phi  \right ]  
&.&&
\end{array}
\label{eq:lind2}
\end{equation}
where
$\widehat{W}_{\mu\nu} \equiv  W^j_{\mu\nu} \frac{\sigma^j}{2}$
is the $SU(2)_L$ field strength while $B_{\mu\nu}$ stands for the
$U(1)_Y$ one.\smallskip

In addition to the above operators, there are also genuine QGC ones
that contain just field strengths:
\begin{equation}
\begin{array} {lcl}
 {\cal O}_{T,0} =   \hbox{Tr}\left [ \widehat{W}_{\mu\nu} \widehat{W}^{\mu\nu} \right ]
\times   \hbox{Tr}\left [ \widehat{W}_{\alpha\beta} \widehat{W}^{\alpha\beta} \right ]
&, &
 {\cal O}_{T,1} =   \hbox{Tr}\left [ \widehat{W}_{\alpha\nu} \widehat{W}^{\mu\beta} \right ]
\times   \hbox{Tr}\left [ \widehat{W}_{\mu\beta} \widehat{W}^{\alpha\nu} \right ]
\\
 {\cal O}_{T,2} =   \hbox{Tr}\left [ \widehat{W}_{\alpha\mu} \widehat{W}^{\mu\beta} \right ]
\times   \hbox{Tr}\left [ \widehat{W}_{\beta\nu} \widehat{W}^{\nu\alpha} \right ]
&,&
         {\cal O}_{T,3}
         =  \hbox{Tr}\left [ \widehat{W}_{\mu\nu} \widehat{W}_{\alpha\beta} \right ]
\times   \hbox{Tr}\left [ \widehat{W}^{\alpha\nu} \widehat{W}^{\mu\beta} \right ]
 \\
{\cal O}_{T,4} =   \hbox{Tr}\left [ \widehat{W}_{\mu\nu} \widehat{W}_{\alpha\beta} \right ]
\times   B^{\alpha\nu} B^{\mu\beta}
 &,&
 {\cal O}_{T,5} =   \hbox{Tr}\left [ \widehat{W}_{\mu\nu} \widehat{W}^{\mu\nu} \right ]
\times   B_{\alpha\beta} B^{\alpha\beta}
\\ 
 {\cal O}_{T,6} =   \hbox{Tr}\left [ \widehat{W}_{\alpha\nu} \widehat{W}^{\mu\beta} \right ]
\times   B_{\mu\beta} B^{\alpha\nu} 
&,&
 {\cal O}_{T,7} =   \hbox{Tr}\left [ \widehat{W}_{\alpha\mu} \widehat{W}^{\mu\beta} \right ]
\times   B_{\beta\nu} B^{\nu\alpha} 
\\ 
 {\cal O}_{T,8} =   B_{\mu\nu} B^{\mu\nu}  B_{\alpha\beta} B^{\alpha\beta}
&,& 
 {\cal O}_{T,9} =  B_{\alpha\mu} B^{\mu\beta}   B_{\beta\nu} B^{\nu\alpha} 
\; . 
\end{array}
\label{eq:lan-t}
\end{equation}

These 20 operators induce all possible modifications to vertices
$VVVV$, $VVVH$ and $VVHH$ ($V=W^\pm$, $Z$ and $A$) that are compatible
with electric charge, $C$ and $P$ conservation; for further details on
the anomalous vertices generated by each dimension-eight operator see
Ref.~\cite{Eboli:2016kko}.\smallskip

\subsubsection{Nonlinear ${\cal O}(p^4)$  realization of the gauge symmetry}
\label{sss:nonlin}

In dynamical scenarios, the Higgs boson is a composite state, {\em
  i.e.} it is a pseudo-Nambu-Goldstone boson of an exact global
symmetry.  Therefore, the gauge symmetry of the low-energy effective
lagrangian is realized nonlinearly~ \cite{Weinberg:1978kz,
  Feruglio:1992wf, Appelquist:1980vg, Longhitano:1980tm} and the
effective lagrangian is a derivative expansion.  In this case, the
effective Lagrangian is written in terms of the SM fermions and gauge
bosons and of the physical Higgs $h$.  The building block at low
energies is a dimensionless unitary matrix transforming as a
bi-doublet of the global symmetry $SU(2)_L \otimes SU(2)_R$:
\begin{equation}
\UH(x)=e^{i\sigma_a \pi^a(x)/v}\; , \qquad \qquad  \UH(x) \rightarrow L\, \UH(x) R^\dagger\;,
\end{equation}
where $L$, $R$ denote $SU(2)_{L,R}$ global transformations,
respectively and $\pi^a$ are the Goldstone bosons.  Its covariant
derivative is given by
\begin{equation}
\DLR_\mu \UH(x) \equiv \partial_\mu \UH(x) +ig \frac{\sigma^j}{2}
W^i_{\mu}(x)\UH(x) - \frac{ig'}{2}  B_\mu(x) \UH(x)\sigma_3 \; .
\end{equation}
From this basic element it is possible to construct the vector chiral
field
\begin{equation}
V_\mu \equiv   \left(\DLR_\mu\UH\right)\UH^\dagger \;,
\end{equation}
and the scalar chiral field $T\equiv\UH\sigma_3\UH^\dag$.  For further
details see Ref.~\cite{Eboli:2016kko}. \smallskip

The lowest operators respecting $C$ and $P$ and exhibiting genuine QGC
are of order $p^4$, that in the notation of Refs.~\cite{Alonso:2012px,
  Brivio:2013pma} are
\begin{equation}
\begin{array}{ll}
{\cal P}_6 = \Tr [ V^\mu V_\mu] \Tr[V^\nu V_\nu] {\cal F}_6(h)
  \;\;\hbox{  ,  }\;\; 
&
{\cal P}_{11} = \Tr [ V^\mu V^\nu] \Tr[V_\mu V_\nu] {\cal F}_{11}(h)
\;,  
\label{eq:p41}
\end{array}
\end{equation}
that respect the $SU(2)_c$ custodial symmetry and
\begin{equation}
\begin{array}{ll}
{\cal P}_{23} = \Tr [ V^\mu V_\mu] (\Tr[ T V_\nu])^2 {\cal F}_{23}(h)
  \;\;\hbox{  ,  }\;\; 
&
  {\cal P}_{24} = \Tr [ V^\mu V^\nu] \Tr[ T V_\mu] \Tr[T V_\nu]
{\cal F}_{24}(h)
  \\
  & \\
{\cal P}_{26} = ( \Tr[ T V_\mu] \Tr[T V_\nu] )^2{\cal F}_{26}(h)
&
\; ,
\label{eq:p42}
\end{array}
\end{equation}
which violate $SU(2)_c$.  ${\cal F}_i(h)$ are generic functions
parametrizing the chiral-symmetry breaking interactions of $h$. As we
are looking for operators whose lowest order vertex contains four gauge
bosons, we take ${\cal F}_i=1$.  So, the most general Lagrangian at
${\cal O}(p^4)$ for genuine QGC is
\begin{equation}
{\cal L}^{p=4}_{QGC}=\sum_{i=6,11,23,24,26} f_{P,i} {\cal P}_i \;.
\end{equation} 
It is interesting to notice that the above nonlinear operators do not
contain photons.  \smallskip

\subsection{Partial-wave unitarity}
\label{sec:unit}

In the two-to-two scattering of electroweak gauge bosons ($V$)
\begin{equation}
{V_1}_{\lambda_1}{V_2 }_{\lambda_2} \to {V_3}_{\lambda_3}{V_4}_{\lambda_4}
\end{equation}
the corresponding helicity amplitude can be expanded in partial waves
in the center--of-mass system as \cite{Jacob:1959at}
\begin{equation}
\mathcal{M}
({V_1}_{\lambda_1}{V_2 }_{\lambda_2} \to {V_3}_{\lambda_3}{V_4}_{\lambda_4}) 
=16 \pi \sum_J
\left (2 J+1 \right)~ 
\sqrt{1+\delta_{{V_1}_{\lambda_1}}^{{V_2}_{\lambda_2}}}
\sqrt{1+\delta_{{V_3}_{\lambda_3}}^{{V_4}_{\lambda_4}}}
d_{\lambda\mu}^{J}(\theta) ~e^{i M \varphi}
~ T^J({V_1}_{\lambda_1}{V_2 }_{\lambda_2} \to {V_3}_{\lambda_3}{V_4}_{\lambda_4}) 
\;\;,
\label{eq:helamp}
\end{equation}
where $\lambda=\lambda_1-\lambda_2$, $\mu=\lambda_3-\lambda_4$,
$M = \lambda_1 - \lambda_2 - \lambda_3 + \lambda_4$, and $\theta$
($\varphi$) is the polar (azimuth) scattering angle. $d$ is the usual
Wigner rotation matrix. For processes where we substitute a vector
boson by a Higgs, this expression can be used by setting the
correspondent $\lambda$ to zero.\smallskip

Partial-wave unitarity for a given elastic channel requires that
\begin{equation}
|T^J({V_1}_{\lambda_1}{V_2 }_{\lambda_2} \to {V_1}_{\lambda_1}{V_2}_{\lambda_2}) 
| \le 1 \;\;,
\label{eq:unitcond}
\end{equation}
where we considered the limit $s\gg (M_{V_1}+M_{V_2})^2$.  More
stringent bounds can be obtained by diagonalizing $T^J$ in the
particle and helicity space and then applying the condition in
Eq.~(\ref{eq:unitcond}) to each of the eigenvalues. \smallskip

In our analysis we evaluated $T^0$ and $T^1$ amplitude matrices in
particle and parameter space as a function of the Wilson coefficients
of the dimension-eight operators and the nonlinear ones. These
matrices are formed with the $s$-divergent parts of the amplitudes
corresponding to all combinations of gauge boson and Higgs pairs with
a given total charge $Q=2,1,0$ with possible projections on a given
partial wave $J$ which are:
\begin{equation}
\begin{array}{c | ccccccccccc r}
(Q,J)			&{\rm States}	&&&&&&&&&& \rm{Total} \\
\hline
\hline
(2,0) 		&	W^+_\pm W^+_\pm		&W^+_0W^+_0
&&&&&&&&&&3\\[+0.1cm]
\hline
(2,1) 		&W^+_\pm W^+_0 	&W^+_0W^+_\pm
&&&&&&&&&& 4	\\[+0.1cm]
\hline
(1,0) 		&	W^+_\pm Z_\pm		&W^+_0Z_0		
		&W^+_\pm \gamma_\pm		&W^+_0H		
&&&&&&&&6\\[+0.1cm]
\hline
(1,1) 		&	W^+_0Z_0		&W^+_\pm Z_0	&W^+_0Z_\pm
	&W^+_\pm Z_\pm	&	W^+_0\gamma_\pm	&W^+_\pm\gamma_\pm
		&W^+_0H					&W^+_\pm H	
&&&&14\\ [+0.1cm]
\hline
(0,0) 		&	W^+_\pm W^-_\pm		&W^+_0W^-_0		
		&Z_\pm Z_\pm	&Z_0Z_0	  	&	Z_\pm\gamma_\pm		
&\gamma_\pm\gamma_\pm	&Z_0H	&HH		
&&&&12\\[+0.1cm]
\hline
(0,1)		&	W^+_0W^-_0			&W^+_\pm W^-_0
		&W^+_0W^-_\pm	&W^+_\pm W^-_\pm	&	Z_\pm Z_0
		&Z_0Z_\pm 				&Z_\pm\gamma_\pm	&Z_0\gamma_\pm	
   &	Z_0H				&Z_\pm H	&\gamma_\pm H	
& 20\\[+0.1cm]
\end{array}
\label{eq:blocks}
\end{equation}
where upper indices indicate charge and lower indices helicity.  We
also display in Eq.~(\ref{eq:blocks}) the dimensionality of the particle
and helicity matrix for each independent $(Q \, , J)$ channel.  Parity
conservation at tree level leads to the reduction of number of
independent helicity amplitudes once we take into account the relation
\begin{equation}
 T^J({V_1}_{\lambda_1}{V_2 }_{\lambda_2} \to {V_3}_{\lambda_3}{V_4}_{\lambda_4})
=(-1)^{\lambda_1-\lambda_2-\lambda_3+\lambda_4}
  T^J({V_1}_{-\lambda_1}{V_2 }_{-\lambda_2} \to {V_3}_{-\lambda_3}{V_4}_{-\lambda_4}) 
\;\;.
\end{equation}
Furthermore, time-reversal invariance also reduces the number of
helicity amplitudes that need to be evaluated. \smallskip


At this point, we would like to point out that there are further
  dimension-eight and $p^4$ nonlinear operators that can contribute to
  the processes listed in Eq.~(\ref{eq:blocks}). For instance, the
  operators
\begin{equation}
B^{\mu\nu} ~\Phi^\dagger \Phi ~(D_\mu \Phi)^\dagger D^\mu\Phi
\;\;\;\hbox{ and }\;\;\; \hbox{Tr} [ D_\mu U U^\dagger D_ \nu U
U^\dagger] ~\partial^\mu {\cal F}_8 \partial^\nu {\cal F}^\prime_8
\end{equation}
contribute to $ VV \to HH$; for a complete list see
Refs.~\cite{Li:2020gnx, Murphy:2020rsh, Brivio:2014pfa}. Here, we
focus on effective operators leading to genuine quartic gauge
interactions and we do not attempt to perform a full analysis for
dimension-eight ($p^4$) operators. \smallskip


As an illustration, let us study the $Q=2$ and $J=0$ channel.  The
leading term in the center-of-mass energy $\sqrt{s}$ of the unitarity
violating amplitudes is ${\cal O}(s^2)$ as expected from a naive dimensional
analysis\footnote{Since genuine QGC do not have a TGC counterpart,
  gauge invariance does not lead to the cancellation of the $s^2$
  terms~\cite{Csaki:2003dt} in contrast to what happens for
  dimension-six QGC~\cite{Corbett:2014ora}.}.  Working in the basis
$\left(W^+_{+}W^+_{+} ,W^+_{0}W^+_{0} , W^+_{-}W^+_{-}\right)$, the
$3 \times 3$ matrix in helicity particle space reads
\begin{equation}
\frac{1}{96\pi} \frac{s^2}{\Lambda^4}\left(
\begin{array} {ccc}
6 f_{T_1} + 3 f_{T_2} + 3 f_{T_3} 
& 0 
& 
4 f_{T_0} + 8 f_{T_1}  + f_{T_2} + 3 f_{T_3}
\\
0
& 3 f_{S_0} + f_{S_1} + f_{S_2}
&0
\\
4 f_{T_0} + 8 f_{T_1}  + f_{T_2} + 3 f_{T_3}
&0
&6 f_{T_1} + 3 f_{T_2} + 3 f_{T_3}
\\
\end{array}
\right)  \;\;.
\label{eq:example}
\end{equation}   
The strongest unitarity limits from this channel come from the 
eigenvalues of the above matrix:
\begin{align}
\left| \Frac{3 f_{S_0} + f_{S_1} + f_{S_2}}{\Lambda^4} s^2    \right|
&\le  96 \pi \; , &
\left| \Frac{2 f_{T_0} + f_{T_1} - f_{T_2}}{\Lambda^4} s^2    \right|
&\le  48 \pi \;,&
\left| \Frac{2 f_{T_0} + 7 f_{T_1} + 2 f_{T_2}+ 3 f_{T_3} }{\Lambda^4} s^2    \right|
&\le  48 \pi \;.
\end{align}
Clearly this allows to constrain the coefficients only under the
assumption of no cancellations between the different coefficients. So
in order to obtain the most stringent bounds on the full set of
coefficients we diagonalize the six $T^J$ matrices and impose the
constraint Eq.~(\ref{eq:unitcond}) on each of their eigenvalues.
\smallskip

\section{Results}
\label{sec:results}

We start our analysis studying the operators that contain four
covariant derivatives of the Higgs field, which are given in
Eq.~(\ref{eq:dphi}). The strongest unitarity limits for these
operators originate from the $J=0$ partial wave. When we diagonalize
the helicity-particle matrices for the three charges ($Q=0,1,2$) we
obtain three distinct nonvanishing eigenvalues given by
\begin{align}
   \Frac{s^2}{96\,\pi}\left(
  \Frac{3\,f_{S,0}\,+\,f_{S,1}\,+\,f_{S,2}}{\Lambda^4}\right)\;,
  &&
   \Frac{s^2}{96\,\pi}\left(\Frac{f_{S,0}\,+\,f_{S,1}\,+\,3\,f_{S,2}}{\Lambda^4}\right) \;, &&
   \Frac{s^2}{96\,\pi}\left(\Frac{3\,f_{S,0}\,+\,7\,f_{S,1}\,+\,5\,f_{S,2}}{\Lambda^4}\right) \;,
\label{eq:ev-ls}
\end{align}
where we kept only the leading term in the center-of-mass energy. \smallskip

These eigenvalues allow us to obtain limits on the three Wilson
coefficients $f_{S,j}/\Lambda^4$.  In order to explore the dependence
of the bounds on the possible relations among operators imposed by
specific forms of the ultraviolet physics, we consider two scenarios:
in the first one, we assume that only one Wilson coefficient is
nonvanishing. The second case assumes that all Wilson coefficients of
the subset of operators considered are nonvanishing and we look for
the possible largest values of each coefficient in the unitarity
region. Notice that in this second scenario the limits for all
couplings cannot be achieved simultaneously as they are simply extreme
points in the three-dimension
($f_{S,0}/\Lambda^4, f_{S,1}/\Lambda^4, f_{S,2}/\Lambda^4$) region
delimited by Eqs.~(\ref{eq:unitcond}) and (\ref{eq:ev-ls}).  We
present in Table~\ref{tab:ls} the bounds on $f_{S,j}/\Lambda^4$ for
these two scenarios. Expectedly, the limits in the second case are
weaker since the undisplayed Wilson coefficients can be adjusted to
mitigate unitarity violation but as explicitly shown this is only an
effect of ${\cal O} (1.5$--$4)$. \smallskip

\begin{table}
\begin{tabular}{|c|c|c|c|c|}
\hline
 & \multicolumn{4}{c|}{Bound}\\ \cline{2-5}
Wilson & \multicolumn{2}{c|}{1 operator} & \multicolumn{2}{c|}
{all 3 operators}
\\\cline{2-5}
Coefficient&  & 
For $\sqrt{s}<1.5\, (3)$ TeV &
& For $\sqrt{s}<1.5\, (3)$ TeV
\\\hline
&&&&\\[-0.2cm]
$\left|\Frac{f_{S,0}}{\Lambda^4}\right|$
& $32 \,\pi\,s^{-2}$ 
& 20 (1.2)  TeV$^{-4}$
  & $48\,\pi\,s^{-2}$
& 30 (1.9) TeV$^{-4}$
  \\[+0.2cm]
  $\left|\Frac{f_{S,1}}{\Lambda^4}\right|$&
  $\Frac{96}{7} \,\pi\,s^{-2}$ &
8.5 (0.53) TeV$^{-4}$
  & 
$\Frac{288\, \pi}{5}\,s^{-2}$
& 35 (2.2) TeV$^{-4}$
  \\[+0.2cm]
  $\left|\Frac{f_{S,2}}{\Lambda^4}\right|$&
  $\Frac{96}{5} \,\pi \,s^{-2}$ &
  8.5 (0.53) TeV$^{-4}$
  &
  $\Frac{288\, \pi}{5}\,s^{-2}$
&  35 (2.2) TeV$^{-4}$
  \\
  \hline
\end{tabular}
\caption{Unitarity constraints on the Wilson coefficients of the
  ${\cal O}_{S,j}$ operators (Eq.~\ref{eq:dphi}) when just one
  coefficient is nonvanishing (second and third column), as well as,
  when all coefficients are included (last two columns).  For
  convenience, in the third and fifth columns, we give the numerical
  value of the bounds for maximal subprocess center-of-mass energy of
  1.5 and 3 TeV.}
\label{tab:ls}
\end{table}

Next, we focus on the unitarity constraints on the seven operators
${\cal O}_{Mj}$ from their leading contributions (${\cal O}(s^2)$) to
the scattering amplitudes. For these operators, the analysis of the
$J=0$ partial wave for the three charges yields two independent
nonvanishing eigenvalues
\begin{align}
  \frac{s^2}{64 \,\pi\,\Lambda^4}
\left(2 f_{M,4}+ f_{M,5}\right)\;, &&
   \frac{s^2}{256 \,\pi\,\Lambda^4}  \sqrt{32 \,
     \left(-4 f_{M,2}+f_{M,3}\right)^2\,+\, 6\, \left(8 f_{M,0}-2
                                      f_{M,1}+f_{M,7} \right)^2}\; .
\label{eq:ev-lm1}   
\end{align}  
They allow for constraining the Wilson coefficients when considering
only one operator at a time.  Nevertheless, they are not enough to
bound all coefficients in the most general scenario with several
nonvanishing operators entering the amplitudes simultaneously.
Consequently, to obtain the limits from the leading ${\cal O}(s^2)$
contribution in this case one must also consider the bounds from the
$J=1$ partial-wave unitarity. In so doing, we find seven additional
independent nonzero eigenvalues in the $Q=2,1$ helicity-particle
matrices
\begin{align}
    \Frac{s^2}{1536 \,\pi\,\Lambda^4} C_{1,2,3}\; & &
    \Frac{s^2}{6144 \,\pi\,\Lambda^4} \left ( C_4\pm \sqrt{{C_5}^2
      +{C_6}^2} \right)\;, & &
    \Frac{s^2}{6144 \,\pi\,\Lambda^4} \left ( C_{7}\pm \sqrt{{C_{8}}^2+{C_{9}}^2} \right)\;,
\label{eq:ev-lm2}
\end{align}
with
\begin{align*}
  C_1&= 12f_{M,0}+ 5f_{M,1}\;,&
  C_2&= 12 f_{M,0}-11 f_{M,1}+8f_{M,7}\;,&
  C_3&= 4 f_{M,7}\;,\\
  C_{4,5}&=\pm(24 f_{M,0}+10 f_{M,1} -15f_{M,7})+48 f_{M,2}+20 f_{M,3}
  \;, &
  C_6&=4\sqrt{3}(6f_{M,4}-5 f_{M,5})\;, & &\\
  C_{7,8}&=\pm(-24 f_{M,0}+22 f_{M,1}-f_{M,7})-48 f_{M,2}+44 f_{M,3}
  \;,&
  C_{9}&=4\sqrt{3}(6f_{M,4}+11 f_{M,5})\;. & &
\end{align*}
Altogether, the total number of unitarity constraints originating from
the unitarity condition Eq.~(\ref{eq:unitcond}) with
Eqs.~(\ref{eq:ev-lm1}) and (\ref{eq:ev-lm2}), allow for independently
bounding each of the $f_{M,j}/\Lambda^4$ Wilson coefficients even when
all seven are considered simultaneously. In fact, due to the algebraic
structure of the eigenvalues, it is technically possible to solve
analytically the system of 9 constraints in the seven dimensional
parameter space. \smallskip

We present in Table~\ref{tab:lm} the unitarity bounds on the Wilson
coefficients of the operators ${\cal O}_{M,j}$ for the two scenarios
described above. Even though the $J=1$ partial waves have to be
invoked to obtained bounds in the full seven dimensional parameter
space and the limits of higher angular momentum amplitudes are weaker,
it is interesting that the constraints on the Wilson coefficients do
not degrade substantially and become ${\cal O} (3$--$4)$ weaker than
those obtained from the $J=0$ partial waves under the assumption of
only one nonvanishing coefficient. \smallskip

\begin{table}
\begin{tabular}{|c|c|c|c|c|}
\hline
 & \multicolumn{4}{c|}{Bound}\\ \cline{2-5}
Wilson & \multicolumn{2}{c|}{1 operator} & \multicolumn{2}{c|}
{all  7 operators}
\\\cline{2-5}
Coefficient&  & 
For $\sqrt{s}<1.5\, (3)$ TeV &
& For $\sqrt{s}<1.5\, (3)$ TeV
\\\hline
&&&&\\[-0.2cm]
$\left | \Frac{f_{M,0}}{\Lambda^4} \right|$ &
$\Frac{32}{\sqrt{6}} \,\pi \,s^{-2}$
& 
8.1 (0.5) TeV$^{-4}$
&
{$ \Frac{2}{3}(72+5\sqrt{6})\,\pi)\,s^{-2}$ }
&
35 (2.1)    TeV$^{-4}$
\\[+0.2cm]
$\left | \Frac{f_{M,1}}{\Lambda^4} \right|$
& $\Frac{128}{\sqrt{6}} \,\pi \,s^{-2}$
&
32 (2) TeV$^{-4}$
& 
  {$ 8\left(24+\Frac{\sqrt{6}}{5}\right)\,\pi\,s^{-2}$}
  &
122 (7.6)  TeV$^{-4}$
\\[+0.2cm]
$\left | \Frac{f_{M,2}}{\Lambda^4} \right|$
&$\Frac{16}{\sqrt{2}} \,\pi\,s^{-2}$
&
7 (0.44) TeV$^{-4}$
&
  {$ (24+5\sqrt{2})\,\pi\,s^{-2}$}
  &
  20 (1.3) TeV$^{-4}$
\\[+0.2cm]
$\left | \Frac{f_{M,3}}{\Lambda^4} \right|$
&$\Frac{64}{\sqrt{2}} \,\pi\,s^{-2}$
&
28 (1.7) TeV$^{-4}$
& {$ 96\,\pi\,s^{-2}$}
&
60 (3.7) TeV$^{-4}$
\\[+0.2cm]
$\left | \Frac{f_{M,4}}{\Lambda^4} \right|$
&$32 \,\pi\,s^{-2}$
&
20 (1.2) TeV$^{-4}$
& $4(5+8\sqrt{3})\,\pi\,s^{-2}$
&
58 (3.6) TeV$^{-4}$
\\[+0.2cm]
$\left | \Frac{f_{M,5}}{\Lambda^4} \right|$
&$64 \,\pi\,s^{-2}$
&
40 (2.5) TeV$^{-4}$
& {$ 64\,\sqrt{3}\,\pi\,s^{-2}$}
&
69 (4.3) TeV$^{-4}$
\\[+0.2cm]
$\left | \Frac{f_{M,7}}{\Lambda^4} \right|$
&$\Frac{256}{\sqrt{6}} \,\pi\,s^{-2}$
&
65 (4.0) TeV$^{-4}$
& {$ \Frac{64}{5}(24+\sqrt{6})\pi\,s^{-2}$}
&
210 (13) TeV$^{-4}$
\\
\hline
\end{tabular}
\caption{Same as Table~\ref{tab:ls} but for the operators 
${\cal O}_{M,j}$ (Eq.~\eqref{eq:lind2}). }
\label{tab:lm}
\end{table}

The third class of dimension-eight operators exhibits only field
strength tensors and it contains ten independent operators given in
Eq.~(\ref{eq:lan-t}). Considering only leading contributions to the
scattering amplitudes that grow as $s^2$, the diagonalization of the
$J=0$ helicity-particle matrices for the three $Q$ channels leads to
eight distinct eigenvalues
\begin{align}
\frac{s^2 }{384 \,\pi\,\Lambda^4} D_{1,2,3,4} \;, &&
 \frac{s^2}{384 \,\pi\,\Lambda^4}  \left(D_5\pm\sqrt{D_6^2+D_7^2}\right)\;, &&
\frac{s^2}{384 \,\pi\,\Lambda^4}
\left(D_8\pm\sqrt{D_9^2+D_{10}^2}\right)\; ,
\label{eq:ev-lt1}
\end{align}
where 
\begin{align*}
D_1&=  8(2 f_{T,0} +  f_{T,1} - f_{T,2})\;, &
D_2&= 8(2 f_{T,0} + 7 f_{T,1} + 2 f_{T,2}+ 3 f_{T_3} ) \;, \\
D_3&= 8(2 f_{T,5} + f_{T,6} - f_{T,7}) \;, &
D_4&= 8(2 f_{T,5} + 7 f_{T,6} + 2 f_{T,7}+ 3 f_{T_4} ) \;, \\
D_{5,6}&= 4 (2 f_{T,0} + 4 f_{T,1} - f_{T,2} \pm( 8 f_{T,8} - 4
       f_{T,9} - 3 f_{T_3} ))  \;, &
D_7&= 8 \sqrt{3} (f_{T,6}- f_{T_4} ) \;,\\
D_{8,9}&= 80 f_{T,0} + 40 f_{T,1} + 26 f_{T,2} \pm( 128 f_{T,8} + 56
       f_{T,9}+ 18 f_{T_3} ) \;, &
D_{10}&= 4 \sqrt{3} (12 f_{T,5} + 2 f_{T,6} + 3 f_{T,7}+  f_{T_4} ) 
        \;.
\end{align*}
In this case the total number of unitarity constraints originating
from the unitarity condition Eq.~(\ref{eq:unitcond}), together with
the $J=0$ eigenvalues in Eq.~(\ref{eq:ev-lt1}) allow for independently
bounding each of the $f_{T,j}/\Lambda^4$ Wilson coefficients even when
the ten are considered simultaneously. And again, it is technically
possible to solve analytically the system of 8 constraints in the
ten-dimensional parameter space. The bounds emanating from the $J=1$
partial wave are weaker than the ones from the $J=0$ one, therefore,
we neglected them in this analysis. \smallskip

We list in Table~\ref{tab:lt} the corresponding bounds for the
$f_{T,j}/\Lambda^4$ coefficients assuming the two scenarios described
above. Comparing the results in Tables~\ref{tab:ls}--\ref{tab:lt} we
learn that the Wilson coefficients of the operators ${\cal O}_{T,j}$
are subject to stronger unitarity bounds than the other QGC
classes. Moreover, for only one nonvanishing $f_{T,j}/\Lambda^4 =1$
TeV$^{-4}$, unitarity is not violated for subprocess center-of-mass
energies smaller than 1.5--2.8 TeV depending on the anomalous QGC.
\smallskip

\begin{table}
\begin{tabular}{|c|c|c|c|c|}
\hline
 & \multicolumn{4}{c|}{Bound}\\ \cline{2-5}
Wilson & \multicolumn{2}{c|}{1 operator} & \multicolumn{2}{c|}
{all 10 operators}
\\\cline{2-5}
Coefficient&  & 
For $\sqrt{s}<1.5\, (3)$ TeV &
& For $\sqrt{s}<1.5\, (3)$ TeV
\\\hline
&&&&\\[-0.2cm]
  $\left|\Frac{f_{T,0}}{\Lambda^4}   \right|$ 
& $\Frac{12}{5} \,\pi\,s^{-2}$
& 1.5 (0.093) TeV$^{-4}$
& {$\Frac{40}{3} \,\pi\,s^{-2}$}
& {8.3 (0.52) TeV$^{-4}$}
\\[+0.2cm]
  $\left|\Frac{f_{T,1}}{\Lambda^4}   \right|$ 
& $\Frac{24}{5} \,\pi\,s^{-2}$
& 3.0 (0.19) TeV$^{-4}$
& {$\Frac{176}{15} \,\pi\,s^{-2}$}
& {7.3 (0.46) TeV$^{-4}$}
\\[+0.2cm]
  $\left|\Frac{f_{T,2}}{\Lambda^4}   \right|$ 
&$\Frac{96}{13} \,\pi\,s^{-2}$
& 4.6 (0.29) TeV$^{-4}$
& {$ \Frac{208}{5} \,\pi\,s^{-2}$}
& {26 (1.6) TeV$^{-4}$}
\\[+0.2cm]
  {$\left|\Frac{f_{T,3}}{\Lambda^4}   \right|$} 
  &{$\Frac{32}{3} \,\pi\,s^{-2}$}
  & {6.6 (0.41) TeV$^{-4}$}
  &  {$\Frac{176}{5} \,\pi\,s^{-2}$}
  & {22 (1.4) TeV$^{-4}$}
  \\[+0.2cm]
  {$\left|\Frac{f_{T,4}}{\Lambda^4}   \right|$} 
  &{$16 \,\pi\,s^{-2}$}
  &{9.9 (0.62) TeV$^{-4}$}
  &  {$8\left(\Frac{6+ 8\sqrt{3}}{5}\right) \,\pi\,s^{-2}$}
  & {20 (1.2) TeV$^{-4}$}
  \\[+0.2cm]
  $\left|\Frac{f_{T,5}}{\Lambda^4}   \right|$ 
& $\Frac{8}{\sqrt{3}} \,\pi\,s^{-2}$
& 2.9 (0.18) TeV$^{-4}$
& {$8\left(\Frac{3+ \sqrt{3}}{3}\right) \,\pi\,s^{-2}$}
& {7.8 (0.49) TeV$^{-4}$}
\\[+0.2cm]
  $\left|\Frac{f_{T,6}}{\Lambda^4}   \right|$ 
&$\Frac{48}{7} \,\pi\,s^{-2}$
& 4.2 (0.27) TeV$^{-4}$
& {$16\left(\Frac{9+5\sqrt{3}}{15}\right) \,\pi\,s^{-2}$}
& {12 (0.73) TeV$^{-4}$}
\\[+0.2cm]
  $\left|\Frac{f_{T,7}}{\Lambda^4}   \right|$ 
&$\Frac{32}{\sqrt{3}} \,\pi\,s^{-2}$
& 11 (0.72) TeV$^{-4}$
& {$32\left(\Frac{5+\sqrt{3}}{5}\right) \,\pi\,s^{-2}$}
& {27 (1.7) TeV$^{-4}$}
\\[+0.2cm]
  $\left|\Frac{f_{T,8}}{\Lambda^4}   \right|$ 
& $\Frac{3}{2} \,\pi\,s^{-2}$
& 0.93 (0.058) TeV$^{-4}$
& {$\Frac{18}{5} \,\pi\,s^{-2}$}
& {2.2 (0.14) TeV$^{-4}$}
\\[+0.2cm]
  $\left|\Frac{f_{T,9}}{\Lambda^4}   \right|$ 
& $\Frac{24}{7} \,\pi\,s^{-2}$
& 2.1 (0.13) TeV$^{-4}$
& {$  8 \,\pi\,s^{-2}$}
&{ 5.0 (0.31) TeV$^{-4}$}
\\
\hline
\end{tabular}
\caption{Same as Table~\ref{tab:ls} but for the operators 
${\cal O}_{T,j}$ (Eq.~\eqref{eq:lan-t}). }
\label{tab:lt}
\end{table}
We end by presenting the unitarity constraints on the ${\cal O}(p^4)$ QCG
given in Eqs.~(\ref{eq:p41}) and (\ref{eq:p42}) that originate in
the nonlinear realization of the gauge symmetry. For this set of
operators the most stringent limits stem from the $J=0$ partial
wave. After diagonalizing the $Q=0,1,2$ channels, we obtain
four non-vanishing  eigenvalues:
\begin{align}
\frac{s^2}{24 v^4 \,\pi\,} F_1 \;, & &
\frac{s^2}{24 v^4 \,\pi\,} F_2 \;, && 
  \frac{s^2}{24 v^4 \,\pi\,}
    \left [ F_3\pm\sqrt{8\,F_4^2+F_5^2} \right] \;, &
\end{align}
where
\begin{align*}
  F_1 &= 4(f_{P,6}+2f_{P,11})\;, &
  F_2 &=  4(f_{P,6}+2f_{P,11}+f_{P,23}+2 f_{P,24}) \;,\\ 
  F_3 &= 13 f_{P,6}+ 11 f_{P,11}+10 f_{P,23}+10 f_{P,24}+20
  f_{P,26} \;, &
  F_4 &= 3 f_{P,6}+ f_{P,11}+3 f_{P,23}+ f_{P,24}\;, \\
  F_5 &= 3 f_{P,6}+ f_{P,11}-10 f_{P,23}-10 f_{P,24}-20 f_{P,26}\;.
\end{align*}
Again, the structure of the four eigenvalues allows for independently
constraining the five $f_{P,i}$ coefficients even when considered all
nonzero simultaneously.  Table~\ref{tab:nonlin} contains the
corresponding bounds on the coefficients. Notice that these results
indicate that the present experimental analyses require the
introduction of a unitarization procedure as the one in
Ref.~\cite{Aaboud:2016ffv}. \smallskip

\begin{table}
\begin{tabular}{|c|c|c|c|c|}
\hline
 & \multicolumn{4}{c|}{Bound}\\ \cline{2-5}
Wilson & \multicolumn{2}{c|}{1 operator} & \multicolumn{2}{c|}
{all 5 operators}
\\\cline{2-5}
Coefficient&  & 
For $\sqrt{s}<1.5\, (3)$ TeV &
& For $\sqrt{s}<1.5\, (3)$ TeV
\\\hline
&&&&\\[-0.2cm]
  $|f_{P,6} |$ & $\Frac{12 \,\pi\, v^4}{11} \,s^{-2}$ &
  2.6 (0.15) $\;\times 10^{-3}$
  &{$ 6 \,\pi\, v^4 \,s^{-2}$}
  & 14 (0.85) $\;\times 10^{-3}$
  \\[+0.2cm]
  $|f_{P,11} |$ & $\Frac{12 \,\pi\, v^4}{7}\,s^{-2}$ &
   3.9 (0.24) $\;\times 10^{-3}$
  &  {$ 6 \,\pi\, v^4\,s^{-2}$}
    & 14 (0.85) $\;\times 10^{-3}$
  \\[+0.2cm]
$|f_{P,23} |$ & $ \Frac{12 \,\pi\, v^4}{5+\sqrt{43} }\,s^{-2}$ &
 2.5    (0.42) $\;\times 10^{-3}$
  &  {$ 12 \Frac{(2+\sqrt{3}) \,\pi\, v^4}{5}\,s^{-2} $}
   &   22 (1.4) $\;\times 10^{-3} $
  \\[+0.2cm]
  $|f_{P,24} |$ & ${  \Frac{12 \,\pi\, v^4} {5+3\sqrt{3}}\,s^{-2} }$ &
    2.9 (0.18) $\;\times 10^{-3 } u$
  &  {${ 12\Frac{(2+\sqrt{3}) \,\pi\, v^4}{5}\,s^{-2} }$}
    &  22 (1.4) $\;\times 10^{-3} $ 
  \\[+0.2cm]
  $|f_{P,26} |$ & $\Frac{3 \,\pi\, v^4}{5}\,s^{-2}$ &
   1.5 (0.09) $\;\times 10^{-3}$
  &  {${  \Frac{12}{5} \,\pi\, v^4\,s^{-2} }$}
           & 5.8 (0.36) $\;\times 10^{-3 }$
             
%
  \\
\hline
\end{tabular}
\caption{Same as Table~\ref{tab:ls} but for the operators 
  in the nonlinear representation of the electroweak symmetry
(Eqs.~\eqref{eq:p41} and \eqref{eq:p42}).}
\label{tab:nonlin}
\end{table}  

\section{Discussion}
\label{sec:discussion}

Exploration of the structure of the quartic couplings of electroweak
gauge bosons is at the forefront of the tests of the SM in general,
and of its mechanism of symmetry breaking in particular.
Parametrizing deviations from the SM predictions in terms of effective
operators is the standard methodology followed in such studies in the
present experimental searches at LHC~\cite{Chatrchyan:2013akv,
  Chatrchyan:2014bza, Aad:2015uqa,
  Khachatryan:2016mud,Khachatryan:2016vif, Aaboud:2016ffv,
  Khachatryan:2017jub, Aaboud:2017tcq, Sirunyan:2017lvq,
  Sirunyan:2017fvv, Sirunyan:2019der, Sirunyan:2020tlu}.
Notwithstanding, the contribution of effective operators leads to
unitarity violation at high energies, and therefore the methodology
must be applied only in the energy regime in which this is not the
case. For the specific case of {\sl genuine} QGC operators, this has
been partially addressed in the literature by studying the bounds
imposed by partial-wave unitarity of gauge-boson scattering in
specific channels and/or waves. \smallskip

In this work we have presented a complete partial-wave analyses of
two-to-two scattering of electroweak gauge bosons and Higgs bosons all
for the charged channels in Eq.~(\ref{eq:blocks}).  We have considered
operator expansions with linear and nonlinear realizations of the
electroweak symmetry. The leading anomalous contribution is
proportional to $s^2$ and we studied the conditions to obtain the most
stringent limits for all couplings.\smallskip

Quantitatively our results are summarize in Tables~\ref{tab:ls},
~\ref{tab:lm},~\ref{tab:lt}, and ~\ref{tab:nonlin}.  In the minimal
scenario with just one nonvanishing QCG Wilson coefficient our
analyses show that the strongest unitarity constraints can be obtained
from the analyses of the $J=0$ partial wave for $Q=0,1,2$.  However,
in more realistic scenarios where more than one QGC operator
contributes, the $J=0$ partial-wave analyses do not lead to the
strongest unitarity bounds for all Wilson-coefficient combinations. In
this case, we must also take into account the $J=1$ partial wave. Once
all waves are considered the bounds on each Wilson coefficient become a
factor of a {\sl few} weaker than in the minimal scenario.\smallskip

\section*{Acknowledgments}

We thank Ilaria Brivio for pointing us that the operators
${\cal O}_{T,3}$ and ${\cal O}_{T,4}$ which were missing in the first
version of this article.
O.J.P.E. is supported in part by Conselho Nacional de Desenvolvimento
Cent\'{\i}fico e Tecnol\'ogico (CNPq) and by Funda\c{c}\~ao de Amparo
\`a Pesquisa do Estado de S\~ao Paulo (FAPESP) grant 2019/04837-9;
E.S.A. thanks FAPESP for its support (grant 2018/16921-1).
M.C.G-G is supported by NSF grant PHY-1620628, by the MINECO grant
FPA2016-76005-C2-1-P, by EU grant FP10 ITN ELUSIVES
(H2020-MSCA-ITN-2015-674896), and by AGAUR (Generalitat de Catalunya)
grant 2017-SGR-929.

%
\bibliography{quarticreferences}
%

\appendix

\section{Helicity Amplitudes}
\label{app:tables}

We present here the list of unitarity violating amplitudes for
all the $2\rightarrow 2$ scattering processes considered in the evaluation
of the unitarity constraints. 

\begin{table}[h]
\begin{tabular} {|c|c|}
\hline
&$\displaystyle\times \frac{s^2}{\Lambda^4}$\\\hline
$W^+W^+\rightarrow W^+ W^+$ &
$ (4\,f_{S,0} + (1 + {X^2})\,(f_{S,1} + f_{S,2}))/8$\\\hline
$W Z\rightarrow WZ$ &
$ ((5 + {2\,X+ X^2})\,(f_{S,0}+f_{S,2})
{+ 2\,(-1 + X)^2}\,f_{S,1} )/16$\\
$W Z\rightarrow WH$ &
$-((-1 + X)\,(3 + X)\,(f_{S,0} - f_{S,2}))/16$\\
$W H\rightarrow WH$ &
$ ((5 + {2\,X+ X^2})\,(f_{S,0}+f_{S,2})
{+ 2\,(-1 + X)^2}\,f_{S,1} )/16$\\
\hline
$W^+W^-\rightarrow W^+ W^-$ &
$({2\,(1+X)^2}\,f_{S,0} {+ (5  -2\,X+X^2)}\,(f_{S,1} + f_{S,2}))/8$\\
$W^+W^-\rightarrow ZZ$ &
$ ((1+X^2)(f_{S,0} + f_{S,2})+ 4\,f_{S,1})/(8\,\sqrt{2})$\\
$W^+W^-\rightarrow ZH$ &
$ -(X\,(f_{S,0} - f_{S,2}))/4$\\
$W^+W^-\rightarrow HH$ &
$ -((1+X^2)(f_{S,0} + f_{S,2})+ 4\,f_{S,1})/(8\,\sqrt{2})$\\
$ZZ\rightarrow ZZ$ &
$ ((3 + {X^2})\,(f_{S,0} + f_{S,1} + f_{S,2}))/8$\\
$ZZ\rightarrow HH$ &
$ ((1-X^2)f_{S,0} - 2\,(f_{S,1} + f_{S,2}))/8$\\
$ZH\rightarrow ZH$ & 
$ (4\,(1 + X)\,f_{S,0}{ + (1 - X)^2}\,(f_{S,1} + f_{S,2}))/8$\\
$HH\rightarrow HH$ & 
$ (3 + {X^2})\,(f_{S,0} + f_{S,1} + f_{S,2})/8$\\ \hline
\end{tabular}
\caption{
  Unitarity violating (growing as $s^2$) terms of the scattering amplitudes
 $\mathcal{M}
({V_1}_{\lambda_1}{V_2 }_{\lambda_2} \to {V_3}_{\lambda_3}{V_4}_{\lambda_4})$ 
for longitudinal gauge bosons generated by the operators
that contain four
covariant derivatives of the Higgs field, 
(Eq.~(\ref{eq:dphi})). $X\equiv\cos\theta$ and 
the overall factor extracted
from all amplitudes is given at the top of the table.}
\label{tab:amps}
\end{table}
\begingroup
  \squeezetable
  \begin{table}
\begin{tabular} {|c||c|l|}
\hline
&\multicolumn{2}{c|}{$\displaystyle\times \frac{s^2}{\Lambda^4}$}\\\hline
$W^+W^+\rightarrow W^+W^+$
& $0+0-$ & $ -X_M\,(-4\,X_M\,f_{M,0} + X_M\,f_{M,1} - 2\,f_{M,7})/64$\\
 & $0+0+$ & $ X_P\,(2\,f_{M,1} - f_{M,7})/32$\\
 & $+00-$ & $ -X_P\,(4\,X_P\,f_{M,0} - X_P\,f_{M,1} + 2\,f_{M,7})/64$\\
 & $+00+$ & $ -X_M\,(2\,f_{M,1} - f_{M,7})/32$\\\hline
$W^+Z\rightarrow W^+Z$ & $00+-$ & $ X_M\,X_P\,f_{M,5}\,s_W/16$\\
 & $00++$ & $ X\,f_{M,7}\,c_W + 2\,(2\,f_{M,4} + f_{M,5})\,s_W/16$\\
 & $0+0-$ & $ X_M^2\,((16\,f_{M,2} - 4\,f_{M,3})\,s_W^2 + 
    (8\,f_{M,0} - 2\,f_{M,1} + f_{M,7})\,c_W^2 - 
    2\,(2\,f_{M,4} + f_{M,5})\,s_{2W})/64$\\
    & $0+0+$ & $ X_P\,(4\,f_{M,3}\,s_W^2 + (2\,f_{M,1} - f_{M,7})\,c_W^2 +     2\,f_{M,5}\,s_{2W})/16$\\
 & $+00-$ & $ X_P\,((-3 + X)\,f_{M,7}\,c_W - 
    2\,X_P\,(2\,f_{M,4} + f_{M,5})\,s_W)/64$\\
 & $+00+$ & $ X_M\,f_{M,5}\,s_W/8$\\\hline
$W^+Z\rightarrow W^+\gamma$
    & $00+-$ & $ -X_M\,X_P\,f_{M,5}\,c_W)/16$\\
 & $00++$ & $ (-2\,(2\,f_{M,4} + f_{M,5})\,c_W + X\,f_{M,7}\,s_W)/16$\\
 & $0+0-$ & $ X_M^2\,(4\,(2\,f_{M,4} + f_{M,5})\,c_{2W} + 
    (8\,f_{M,0} - 2\,f_{M,1} - 16\,f_{M,2} + 4\,f_{M,3} + f_{M,7})\,s_{2W})/128$\\
 & $0+0+$ & $ 
 -X_P\,(4\,f_{M,5}\,c_{2W} + (-2\,f_{M,1} + 4\,f_{M,3} + f_{M,7})\,s_{2W})/32$\\
 & $+00-$ & $ X_P\,(2\,X_P\,(2\,f_{M,4} + f_{M,5})\,c_W + 
    (-3 + X)\,f_{M,7}\,s_W)/64$\\
 & $+00+$ & $ -X_M\,f_{M,5}\,c_W/8$\\\hline
 $W^+Z\rightarrow W^+H$
 & $0+-0$ & $ X_P\,((-3 + X)\,f_{M,7}\,c_W - 
    2\,X_P\,(2\,f_{M,4} + f_{M,5})\,s_W)/64$\\
 & $0++0$ & $ X_M\,f_{M,5}\,s_W/8$\\
 & $+-00$ & $ -X_M\,X_P\,f_{M,5}\,s_W/16$\\
 & $+0-0$ & $ -X_M\,(3 + X)\,f_{M,7}/64$\\
 & $++00$ & $ (-(X\,f_{M,7}\,c_W) - 2\,(2\,f_{M,4} + f_{M,5})\,s_W)/16$\\
\hline
$W^+\gamma\rightarrow W^+\gamma$
& $0+0-$ & $ 
 -X_M^2\,((-8\,f_{M,0} + 2\,f_{M,1} - f_{M,7})\,s_W^2 + (- 16\,f_{M,2} + 4\,f_{M,3})\,
      c_W^2 - 2\,(2\,f_{M,4} + f_{M,5})\,s_{2W})/64$\\
  & $0+0+$ & $ X_P\,((2\,f_{M,1} - f_{M,7})\,s_W^2 + 4\,f_{M,3}\,c_W^2 - 
    2\,f_{M,5}\,s_{2W})/16$\\\hline
    $W^+\gamma\rightarrow W^+H$
    & $0+-0$ & $ X_P\,(2\,X_P\,(2\,f_{M,4} + f_{M,5})\,c_W + 
    (-3 + X)\,f_{M,7}\,s_W)/64$\\
 & $0++0$ & $ -X_M\,f_{M,5}\,c_W/8$\\
 & $+-00$ & $ X_M\,X_P\,f_{M,5}\,c_W/16$\\
 & $++00$ & $ (2\,(2\,f_{M,4} + f_{M,5})\,c_W - X\,f_{M,7}\,s_W)/16$\\\hline
$W^+H\rightarrow W^+H$
    & $+0-0$ & $ X_M^2\,(8\,f_{M,0} - 2\,f_{M,1} + f_{M,7})/64$\\
 & $+0+0$ & $ X_P\,(2\,f_{M,1} - f_{M,7})/16$\\\hline
$W^+W^-\rightarrow W^+W^-$    
    & $00+-$ & $ -X_M\,X_P\,(2\,f_{M,1} - f_{M,7})/32$\\
 & $00++$ & $ (8\,f_{M,0} - 2\,f_{M,1} + f_{M,7} + X\,f_{M,7})/16$\\
 & $0+0-$ & $ -X_M\,(-4\,X_M\,f_{M,0} + f_{M,1} - X\,f_{M,1} + f_{M,7} + X\,f_{M,7})/ 32$\\
        & $0+0+$ & $ X_P\,(2\,f_{M,1} - f_{M,7})/16$\\\hline
$W^+W^-\rightarrow ZZ$
    & $00+-$ & $ X_M\,X_P\,((-2\,f_{M,1} + f_{M,7})\,c_W^2 
     +4\,f_{M,3}\,s^2_W + 2\,f_{M,5}\,s_{2W})/(32\,\sqrt{2})$\\
 & $00++$ & $ ((16\,f_{M,2} - 4\,f_{M,3})\,s_W^2 + (8\,f_{M,0} - 2\,f_{M,1} + f_{M,7})\,
    c_W^2 - 2\,(2\,f_{M,4} + f_{M,5})\,s_{2W})/(16\,\sqrt{2})$\\
 & $0+0-$ & $ -X_M\,((3 + X)\,f_{M,7}\,c_W - 
    2\,X_M\,(2\,f_{M,4} + f_{M,5})\,s_W)/(64\,\sqrt{2})$\\
 & $0+0+$ & $ -X_P\,f_{M,5}\,s_W/(8\,\sqrt{2})$\\
 & $+00-$ & $ 
 -X_P\,((-3 + X)\,f_{M,7}\,c_W + 2\,X_P\,(2\,f_{M,4} + f_{M,5})\,s_W)/
  (64\,\sqrt{2})$\\
 & $+00+$ & $ X_M\,f_{M,5}\,s_W/(8\,\sqrt{2})$\\\hline
 $W^+W^-\rightarrow Z\gamma$
 & $00+-$ & $ 
 X_M\,X_P\,(4\,f_{M,5}\,c_{2W} + (-2\,f_{M,1} + 4\,f_{M,3} + f_{M,7})\,s_{2W})/64$\\
 & $00++$ & $ (2\,(2\,f_{M,4} + f_{M,5})\,c_{2W} + 
   (8\,f_{M,0} - 2\,f_{M,1} - 16\,f_{M,2} + 4\,f_{M,3} + f_{M,7})\,c_W\,s_W)/16$\\
 & $0+0-$ & $ 
 X_M\,(-2\,X_M\,(2\,f_{M,4} + f_{M,5})\,c_W - (3 + X)\,f_{M,7}\,s_W)/64$\\
 & $0+0+$ & $ X_P\,f_{M,5}\,c_W/8$\\
 & $+00-$ & $ X_P\,(2\,X_P\,(2\,f_{M,4} + f_{M,5})\,c_W - 
    (-3 + X)\,f_{M,7}\,s_W)/64$\\
 & $+00+$ & $ -X_M\,f_{M,5}\,c_W/8$\\\hline
 $W^+W^-\rightarrow ZH$
 & $0+-0$ & $ 
 -X_P\,((-3 + X)\,f_{M,7}\,c_W + 2\,X_P\,(2\,f_{M,4} + f_{M,5})\,s_W)/64$\\
 & $0++0$ & $ X_M\,f_{M,5}\,s_W/8$\\
 & $+0-0$ & $ 
 X_M\,((3 + X)\,f_{M,7}\,c_W -2\,X_M\,(2\,f_{M,4} + f_{M,5})\,s_W)/64$\\
 & $+0+0$ & $ X_P\,f_{M,5}\,s_W/8$\\
 & $++00$ & $ X\,f_{M,7}/16$\\\hline
$W^+W^-\rightarrow \gamma\gamma$
 & $00+-$ & $ -X_M\,X_P\,((2\,f_{M,1} - f_{M,7})\,s_W^2 + 4\,f_{M,3} \,c_W^2 - 2\,f_{M,5}\,s_{2W})/(32\,\sqrt{2})$\\
 & $00++$ & $ ((8\,f_{M,0} - 2\,f_{M,1} + f_{M,7})\,s_W^2 - 
   (- 16\,f_{M,2} + 4\,f_{M,3})\,c_W^2 + 
   2\,(2\,f_{M,4} + f_{M,5})\,s_{2W})/(16\,\sqrt{2})$\\\hline
   $W^+W^-\rightarrow \gamma H$ & $0+-0$ &
   $ X_P\,(2\,X_P\,(2\,f_{M,4} + f_{M,5})\,c_W - 
    (-3 + X)\,f_{M,7}\,s_W)/64$\\
 & $0++0$ & $ -X_M\,f_{M,5}\,c_W/8$\\
 & $+0-0$ & $ -X_M\,(-2\,X_M\,(2\,f_{M,4} + f_{M,5})\,c_W - 
(3 + X)\,f_{M,7}\,s_W)/64$\\
& $+0+0$ & $ -X_P\,f_{M,5}\,c_W/8$\\\hline
   $W^+W^-\rightarrow HH$ & $+-00$ &
   $ X_M\,X_P\,(2\,f_{M,1} - f_{M,7})/(32\,\sqrt{2})$\\
 & $++00$ & $ -(8\,f_{M,0} - 2\,f_{M,1} + f_{M,7})/(16\,\sqrt{2})$\\\hline
\end{tabular}
\caption{
Unitarity violating (growing as $s^2$) terms of the scattering amplitudes
 $\mathcal{M}
({V_1}_{\lambda_1}{V_2 }_{\lambda_2} \to {V_3}_{\lambda_3}{V_4}_{\lambda_4})$ 
for the gauge boson helicities given in the second column
generated by the operators that contain two
covariant derivatives of the Higgs field, 
(Eq.~(\ref{eq:lind2})).
$X\equiv\cos\theta$,
  $X_P\equiv 1+\cos\theta$,  $X_M\equiv 1-\cos\theta$,
  $c_W=\cos\theta_W$, $s_W=\sin\theta_W$,
  $c_{2W}=\cos 2\theta_W$, $c_{4W}=\cos 4\theta_W$, and $s_{2W}=\sin 2\theta_W$,
and  the overall factor extracted
from all amplitudes is given at the top of the table.}
\label{tab:ampm1}
  \end{table}
  \endgroup

\begingroup
  \squeezetable
  \begin{table}
\begin{tabular} {|c||c|l|}
\hline
&\multicolumn{2}{c|}{$\displaystyle\times \frac{s^2}{\Lambda^4}$}\\\hline
$ZZ\rightarrow ZZ$
& $00+-$ & $ X_M\,X_P\,(-4\,f_{M,3}\,s_W^2 + (-2\,f_{M,1} + f_{M,7})\,c_W^2 + 
     2\,f_{M,5}\,s_{2W})/64$\\
 & $00++$ & $ ((16\,f_{M,2} - 4\,f_{M,3})\,s_W^2 + (8\,f_{M,0} - 2\,f_{M,1} + f_{M,7})\,
    c_W^2 + 2\,(2\,f_{M,4} + f_{M,5})\,s_{2W})/32$\\
 & $0+0-$ & $ X_M^2\,((16\,f_{M,2} - 4\,f_{M,3})\,s_W^2 + 
    (8\,f_{M,0} - 2\,f_{M,1} + f_{M,7})\,c_W^2 + 
    2\,(2\,f_{M,4} + f_{M,5})\,s_{2W})/128$\\
 & $0+0+$ & $ -X_P\,(-4\,f_{M,3}\,s_W^2 + (-2\,f_{M,1} + f_{M,7})\,c_W^2 + 
     2\,f_{M,5}\,s_{2W})/32$\\
 & $+00-$ & $ 
 -X_P^2\,((16\,f_{M,2} - 4\,f_{M,3})\,s_W^2 + (8\,f_{M,0} - 2\,f_{M,1} + f_{M,7})\,
      c_W^2 + 2\,(2\,f_{M,4} + f_{M,5})\,s_{2W})/128$\\
 & $+00+$ & $ X_M\,(-4\,f_{M,3}\,s_W^2 + (-2\,f_{M,1} + f_{M,7})\,c_W^2 + 
     2\,f_{M,5}\,s_{2W})/32$\\\hline
$ZZ\rightarrow Z\gamma$ & $00+-$ & $ -X_M\,X_P\,(4\,f_{M,5}\,c_{2W} + 
    (2\,f_{M,1} - 4\,f_{M,3} - f_{M,7})\,s_{2W})/(64\,\sqrt{2})$\\
 & $00++$ & $ (-2\,(2\,f_{M,4} + f_{M,5})\,c_{2W} + 
     (8\,f_{M,0} - 2\,f_{M,1} - 16\,f_{M,2} + 4\,f_{M,3} + f_{M,7})\,c_W\,s_W)/(16\,\sqrt{2})$\\
      & $0+0-$ & $ 
 -X_M^2\,(4\,(2\,f_{M,4} + f_{M,5})\,c_{2W} - 
     (8\,f_{M,0} - 2\,f_{M,1} - 16\,f_{M,2} + 4\,f_{M,3} + f_{M,7})\,s_{2W})/(128\,\sqrt{2})$\\
 & $0+0+$ & $ X_P\,(4\,f_{M,5}\,c_{2W} + (2\,f_{M,1} - 4\,f_{M,3} - f_{M,7})\,
     s_{2W})/(32\,\sqrt{2})$\\
 & $+00-$ & $ X_P^2\,(4\,(2\,f_{M,4} + f_{M,5})\,c_{2W} - 
    (8\,f_{M,0} - 2\,f_{M,1} - 16\,f_{M,2} + 4\,f_{M,3} + f_{M,7})\,s_{2W})/(128\,\sqrt{2})$\\
 & $+00+$ & $ -X_M\,(4\,f_{M,5}\,c_{2W} + (2\,f_{M,1} - 4\,f_{M,3} - f_{M,7})\,
     s_{2W})/(32\,\sqrt{2})$\\\hline
     $ZZ\rightarrow \gamma\gamma$ & $00+-$ &
     $ -X_M\,X_P\,((2\,f_{M,1} - f_{M,7})\,s_W^2 + 4\,f_{M,3}\,
     c_W^2 + 2\,f_{M,5}\,s_{2W})/64$\\
 & $00++$ & $ ((8\,f_{M,0} - 2\,f_{M,1} + f_{M,7})\,s_W^2 - 
   (- 16\,f_{M,2} + 4\,f_{M,3})\,c_W^2 - 
   2\,(2\,f_{M,4} + f_{M,5})\,s_{2W})/32$\\\hline
   $ZZ\rightarrow HH$
   & $+-00$ &
   $ -X_M\,X_P\,(-4\,f_{M,3}\,s_W^2 + (-2\,f_{M,1} + f_{M,7})\,c_W^2 + 
    2\,f_{M,5}\,s_{2W})/64$\\
     & $++00$ & $ (4\,(-4\,f_{M,2} + f_{M,3})\,s_W^2 - (8\,f_{M,0} - 2\,f_{M,1}+ f_{M,7})\,
    c_W^2 - 2\,(2\,f_{M,4} + f_{M,5})\,s_{2W})/32$\\\hline
$Z\gamma\rightarrow Z\gamma$ & $0+0-$ & $ 
 -X_M^2\,((-8\,f_{M,0} + 2\,f_{M,1} - f_{M,7})\,s_W^2 + (- 16\,f_{M,2} + 4\,f_{M,3})\,
      c_W^2 + 2\,(2\,f_{M,4} + f_{M,5})\,s_{2W})/64$\\
 & $0+0+$ & $ X_P\,((2\,f_{M,1} - f_{M,7})\,s_W^2 + 4\,f_{M,3} \,c_W^2 + 
    2\,f_{M,5}\,s_{2W})/16$\\\hline
$Z\gamma\rightarrow HH$ & $+-00$ & $ 
 X_M\,X_P\,(4\,f_{M,5}\,c_{2W} + (2\,f_{M,1} - 4\,f_{M,3} - f_{M,7})\,s_{2W})/
  (64\,\sqrt{2})$\\
 & $++00$ & $ (2\,(2\,f_{M,4} + f_{M,5})\,c_{2W} - 
   (8\,f_{M,0} - 2\,f_{M,1} - 16\,f_{M,2} + 4\,f_{M,3} + f_{M,7})\,c_W\,s_W)/(16\,\sqrt{2})$\\\hline
$ZH\rightarrow ZH$ & $+0-0$ & $ X_M^2\,((16\,f_{M,2} - 4\,f_{M,3})\,s_W^2 + 
    (8\,f_{M,0} - 2\,f_{M,1} + f_{M,7})\,c_W^2 + 
    2\,(2\,f_{M,4} + f_{M,5})\,s_{2W}))/64$\\
    & $+0+0$ & $ -X_P\,(-4\,f_{M,3}\,s_W^2 +
    (-2\,f_{M,1} + f_{M,7})\,c_W^2 + 
     2\,f_{M,5}\,s_{2W})/16$\\\hline
$ZH\rightarrow \gamma H$ & $+0-0$ & $ -X_M^2\,(4\,(2\,f_{M,4} + f_{M,5})\,c_{2W} - 
     (8\,f_{M,0} - 2\,f_{M,1} - 16\,f_{M,2} + 4\,f_{M,3} + f_{M,7})\,s_{2W})/128$\\
 & $+0+0$ & $
X_P\,(4\,f_{M,5}\,c_{2W} + (2\,f_{M,1} - 4\,f_{M,3} - f_{M,7})\,
     s_{2W})/32$\\\hline
     $\gamma\gamma\rightarrow H H$
     & $+-00$ & $ -X_M\,X_P\,((-2\,f_{M,1} + f_{M,7})\,s_W^2 - 4\,f_{M,3} \,
     c_W^2 - 2\,f_{M,5}\,s_{2W})/64$\\
 & $++00$ & $ (4\,(-4\,f_{M,2} + f_{M,3})\,c_W^2 + 
   (2\,(2\,f_{M,4} + f_{M,5})\,s_{2W} - (8\,f_{M,0} - 2\,f_{M,1} + f_{M,7})\,s^2_W)/32$\\\hline
$\gamma H\rightarrow \gamma H$ & $+0-0$ & $ 
 -X_M^2\,((-8\,f_{M,0} + 2\,f_{M,1} - f_{M,7})\,s_W^2 + (- 16\,f_{M,2} + 4\,f_{M,3} )\,
      c_W^2 + 2\,(2\,f_{M,4} + f_{M,5})\,s_{2W})/64$\\
      $\gamma H\rightarrow \gamma H$ & $+0+0$ &
      $ X_P\,((2\,f_{M,1} - f_{M,7})\,s_W^2 + 4\,f_{M,3}\,c_W^2 + 
    2\,f_{M,5}\,s_{2W})/16$\\\hline
\end{tabular}
\caption{Continuation of Table ~\ref{tab:ampm1}}
\end{table}
\endgroup  

  \begingroup
  \squeezetable
  \begin{table}
\begin{tabular} {|c||c|l|}
\hline
&\multicolumn{2}{c|}{$\displaystyle\times \frac{s^2}{\Lambda^4}$}\\\hline
  $W^+W^+ \rightarrow W^+W^+$&
$+--+$ & $ X_M^2\,(2\,f_{T,0} + f_{T,1} + f_{T,2}
{+ f_{T,3}}
)/8$ \\
& $+-+-$ & $ X_P^2\,(2\,f_{T,0} + f_{T,1} + f_{T,2}{+ f_{T,3}})/8$ \\
& $++--$ & $
{(4\,(1+X^2)\,f_{T,0} + 2\,(5 + X^2)\,f_{T,1} +
  (1+X^2)\,f_{T,2}) + 4\, f_{T,3}/8 
}$ \\
& $++++$ & $ f_{T,1} + f_{T,2}/2
{+ f_{T,3}/2}
$ \\
\hline
$W^+Z \rightarrow W^+Z$
& $+--+$ & $ X_M^2\,((4\,f_{T,5} + f_{T,7}
{+ f_{T,4}}
)\,s_W^2 + (4\,f_{T,0} + f_{T,2}{+ f_{T,3}} )\,c_W^2))/
  8$ \\
  &$+-+-$& $ X_P^2\,((2\,f_{T,6} + f_{T,7}{+ f_{T,4}})\,s_W^2 + (2\,f_{T,1} + f_{T,2}
  {+ f_{T,3}})\,c_W^2)/
  8$ \\
  &$++--$&$ {(({X_M}^2\,(4 f_{T,5} +f_{T,7})+ 4\,X_P\,f_{T,4}+ 2\,(4+X_P^2)\,f_{T,6} +
   )\,s_W^2   + 
   ({X_M}^2\,(4f_{T,0} +f_{T,2})+ 2\,(4+X_P^2)\,f_{T,1} +4\, X_P\, f_{T,3} )\,c_W^2)/8}$ \\
  &$++++$& $ ((2\,f_{T,6} + f_{T,7}{+ f_{T,4}})\,s_W^2 + (2\,f_{T,1} + f_{T,2}
  {+ f_{T,3}})\,c_W^2)/2$ \\\hline
     $W^+Z \rightarrow W^+\gamma$
  &$+--+$& $ X_M^2\,(4\,f_{T,0} + f_{T,2}{+f_{T,3}-f_{T,4}}  - 4\,f_{T,5} - f_{T,7})\,s_{2W}/16$ \\
&$+-+-$&$ X_P^2\,(2\,f_{T,1} + f_{T,2} {+f_{T,3}-f_{T,4}} - 2\,f_{T,6} - f_{T,7})\,s_{2W}/16$ \\
  &$++--$& ${ (4\,{X_M}^2\,f_{T,0}
    +2\,(4 + X_P^2)\,f_{T,1} + X_M^2 \,f_{T,2} + 4\,{X_P}\,f_{T,3}- 4\,{X_P}\,f_{T,4}
    -4\,X_M^2\,f_{T,5}
    - 2\,(4+X_P)^2\,f_{T,6} - \,X_M^2\,f_{T,7} )\,s_{2W}/16}$
  \\
      &$++++$&$ (2\,f_{T,1} + f_{T,2} {+f_{T,3}-f_{T,4}}- 2\,f_{T,6} - f_{T,7})\,s_{2W}/4$ \\\hline
      $W^+\gamma \rightarrow W^+\gamma$ 
&$+--+$&     
  $ X_M^2\,((4\,f_{T,0} + f_{T,2} {+ f_{T,3}}
  )\,s_W^2 + ({ f_{T,4}}+4\,f_{T,5} + f_{T,7})\,c_W^2)/
  8$ \\
  &$+-+-$& $ X_P^2\,((2\,f_{T,1} + f_{T,2}{+ f_{T,3}})\,s_W^2 +
  ({ f_{T,4}}+2\,f_{T,6} + f_{T,7})\,c_W^2)/
  8$ \\
  &$++--$& ${ (( {X_M}^2\,(4\,f_{T,0} + f_{T,2})+ 2\,(4+X_P^2)\,f_{T,1}
    + 4\,X_P\,f_{T,3}
    )\,s_W^2 
  + (4\,X_P\,f_{T,4}+ X_M^2\,(4\,f_{T,5} + f_{T,7})
     + 2\,(4+X_P^2)\,f_{T,6})\,c_W^2
    ) /8}$
  \\
  &$++++$& $ ((2\,f_{T,1} + f_{T,2}{+ f_{T,3}})\,s_W^2 + (
     { f_{T,4}}+2\,f_{T,6} + f_{T,7})\,c_W^2)/2$ \\
\hline
$W^+W^- \rightarrow W^+W^-$
&$+--+$&
$ X_M^2\,(2\,f_{T,0} + f_{T,1} + f_{T,2}{+ f_{T,3}})/4$ \\
&$+-+-$&$ X_P^2\,(2\,f_{T,1} + f_{T,2}{+ f_{T,3}})/4$ \\
&$++--$&${ (4\,(4 + X_M^2)\,f_{T,0} + 2\,(4 \,X_M + 3\,X_P^2)\,f_{T,1}
  + (4 + X_M^2)\,f_{T,2}+X_P^2\, f_{T,3})/8}$ \\
&$++++$&$ 2\,f_{T,0} + f_{T,1} + f_{T,2}{+ f_{T,3}}$ \\
\hline
$W^+W^- \rightarrow ZZ$ 
&$+--+$& $ X_M^2\,(({ f_{T,4}+}2\,f_{T,6} + f_{T,7})\,s_W^2 + (2\,f_{T,1} + f_{T,2}{+ f_{T,3}})\,c_W^2)/
  (8\,\sqrt{2})$ \\
&$+-+-$& $ X_P^2\,((({ f_{T,4}+}2\,f_{T,6} + f_{T,7})\,s_W^2 + (2\,f_{T,1} + f_{T,2}{+ f_{T,3}})\,c_W^2)/
  (8\,\sqrt{2})$ \\
&$++--$& $ {((-(1-X^2)\,f_{T,4}+8\,f_{T,5} + 2\,(1+X^2)\,f_{T,6} + 2\,f_{T,7})\,s_W^2 + 
   (8\,f_{T,0} + 2\,(1+X^2)\,f_{T,1} + 2\,f_{T,2} -(1-X^2)\,f_{T,3})\,c_W^2)/
  (4\,\sqrt{2})}$ \\
&$++++$& $ (({ f_{T,4}+}4\,f_{T,5} + f_{T,7})\,s_W^2 + (4\,f_{T,0} + f_{T,2}{+ f_{T,3}})\,c_W^2)/(2\,\sqrt{2})$
\\
\hline
$W^+W^- \rightarrow Z\gamma$
&$+--+$&$ X_M^2\,(2\,f_{T,1} + f_{T,2} {+f_{T,3}-f_{T,4}}- 2\,f_{T,6} - f_{T,7})\,s_{2W}/16$ \\
&$+-+-$&$ X_P^2\,(2\,f_{T,1} + f_{T,2}{+f_{T,3}-f_{T,4}} - 2\,f_{T,6} - f_{T,7})\,s_{2W}/16$ \\
&$++--$&${ (8\,f_{T,0} + 2\,(1+X^2)\,f_{T,1} + 2\,f_{T,2}-(1+X^2)\,
  (f_{T,3}-f_{T,4})  - 8\,f_{T,5} - 2\,(1+X^2)\,f_{T,6} - f_{T,7})\,
   s_{2W}/8}$ \\
&$++++$&   $ (4\,f_{T,0} + f_{T,2}{+f_{T,3}-f_{T,4}} - 4\,f_{T,5} - f_{T,7})\,s_{2W}/4$ \\
\hline   
$W^+W^- \rightarrow \gamma\gamma$
&$+--+$&
$ X_M^2\,((2\,f_{T,1} + f_{T,2}{+ f_{T,3}})\,s_W^2 +
({f_{T,4} +}
2\,f_{T,6} + f_{T,7})\,
     c_W^2)/(8\,\sqrt{2})$ \\
&$+-+-$& $ X_P^2\,((2\,f_{T,1} + f_{T,2}{+ f_{T,3}})\,s_W^2 + ({f_{T,4} +}2\,f_{T,6} + f_{T,7})\,c_W^2))/
     (8\,\sqrt{2})$ \\
&$++--$&     
 ${ -((8\,f_{T,0} + 2\,(1+X^2)\,f_{T,1} + 2\,f_{T,2}-(1-X^2)\,f_{T,3})\,s_W^2 + 
   ((1-X^2)\,f_{t,4}-4\,f_{T,5} - 2\,(1+X^2)\,f_{T,6} - 2\,f_{T,7})\,c_W^2)/
     (4\,\sqrt{2})}$ \\
&$++++$&     
     $ ((4\,f_{T,0} + f_{T,2}{+ f_{T,3}})\,s_W^2 + ({f_{T,4} +}4\,f_{T,5} + f_{T,7})\,c_W^2)/(2\,\sqrt{2})$ \\
\hline     
\end{tabular}
\caption{Same as table ~\ref{tab:ampm1} but for 
the operators containing only
  field strength tensors (Eq.~(\ref{eq:lan-t})).}  
\label{tab:ampt1}
  \end{table}
  \endgroup
  
  \begingroup
  \squeezetable
  \begin{table}
\begin{tabular} {|c||c|l|}
\hline
$ZZ \rightarrow ZZ$ 
&$+--+$& $ X_M^2\,((8\,(f_{T,5} + f_{T,6}) + 6\,({f_{T,4}+}f_{T,7}
))\,c_W^2 + 
(4\,(f_{T,0} + f_{T,1}) + 3\,(f_{T,2}{ +f_{T,3}-2\,f_{T,4}}) - 8\,(f_{T,5} +f_{T,6})
- 6\,({f_{T,4}+}f_{T,7})\,c_W^4 $\\&& 
$+ 4\,(4\,f_{T,8} + 3\,f_{T,9})\,s_W^4)/16$ \\
&$+-+-$& $ X_P^2\,((8\,(f_{T,5} + f_{T,6}) + 6\,({f_{T,4}+}f_{T,7}))\,c_W^2 + 
(4\,(f_{T,0} + f_{T,1}) + 3\,(f_{T,2}{ +f_{T,3}-2\,f_{T,4}}) - 8\,(f_{T,5} +f_{T,6})
- 6\,({f_{T,4}+}f_{T,7})\,c_W^4 $\\&& 
$+ 4\,(4\,f_{T,8} + 3\,f_{T,9})\,s_W^4)/16$ \\
&$++--$& $ (3 + {X^2})\,(2\,({f_{T,4}+}
4\,(f_{T,5} + f_{T,6}) + f_{T,7})\,c_W^2 + 
  (4\,(f_{T,0} + f_{T,1}) + (f_{T,2}{ +f_{T,3}-2\,f_{T,4}}- 2\,f_{T,7})
- 8(\,f_{T,5} +f_{T,6}) )\,c_W^4 $\\&&$+ 
4\,(4\,f_{T,8} + f_{T,9})\,s_W^4)/8$ \\
&$++++$& $ ((8\,(f_{T,5} + f_{T,6}) + 6\,({f_{T,4}+}f_{T,7}))\,c_W^2 + 
(4\,(f_{T,0} + f_{T,1}) + 3\,(f_{T,2}{ +f_{T,3}-2\,f_{T,4}}) - 8\,(f_{T,5} +f_{T,6})
- 6\,({f_{T,4}+}f_{T,7})\,c_W^4 $\\&& 
$+ 4\,(4\,f_{T,8} + 3\,f_{T,9})\,s_W^4)/4$ \\
\hline    
$ZZ \rightarrow Z\gamma$
&$+--+$&
$ {X_M^2(-(3\,f_{T,4}+
4\,f_{T,5} + 4\,f_{T,6} + 3\,f_{T,7})\,c_{2W}
- (16\,f_{T,8} + 12\,f_{T,9})\,s^2_W+ 
(4\,f_{T,0} + 4\,f_{T,1} + 3\,f_{T,2} + 3\,f_{T,3}
)\,
     c_W^2)\,s_{2W}/(16\,\sqrt{2})}$ \\
&$+-+-$&
$ {X_P^2(-(3\,f_{T,4}+
4\,f_{T,5} + 4\,f_{T,6} + 3\,f_{T,7})\,c_{2W}
- (16\,f_{T,8} + 12\,f_{T,9})\,s^2_W+ 
(4\,f_{T,0} + 4\,f_{T,1} + 3\,f_{T,2} + 3\,f_{T,3}
)\,
     c_W^2)\,s_{2W}/(16\,\sqrt{2})}$ \\
     &$++--$& $ {(3 + X^2)\,(-(f_{T,4}+4\,f_{T,5} + 4\,f_{T,6} + f_{T,7})c_{2W} - (16\,f_{T,8} + 4\,f_{T,9})\,s_W^2 + 
    (4\,f_{T,0} + 4\,f_{T,1} + f_{T,2}+f_{T,3})\,c_W^2)\,s_{2W}/(8\,\sqrt{2})}$ \\
   &$++++$&
$ {(-(3\,f_{T,4}+
4\,f_{T,5} + 4\,f_{T,6} + 3\,f_{T,7})\,c_{2W}
- (16\,f_{T,8} + 12\,f_{T,9})\,s^2_W+ 
(4\,f_{T,0} + 4\,f_{T,1} + 3\,f_{T,2} + 3\,f_{T,3}
)\,
     c_W^2)\,s_{2W}/(4\,\sqrt{2})}$ \\
\hline   
$ZZ \rightarrow \gamma\gamma$
&$+--+$&
$ -X_M^2\,((-4\,f_{T,0} - 4\,f_{T,1} - 3\,f_{T,2} {-3\,f_{T,3}+2\,f_{T,4}}
  + 8\,f_{T,5} -2 \,f_{T,7}
- 16\,f_{T,8} - 12\,f_{T,9})\,s^2_{2W}
- 4\,({f_{T,4}+} 2\,f_{T,6}+\,f_{T,7})\,c_{2W}^2)/64$ \\
&$+-+-$& 
$ -X_P^2\,((-4\,f_{T,0} - 4\,f_{T,1} - 3\,f_{T,2} {-3\,f_{T,3}+2\,f_{T,4}}
  + 8\,f_{T,5} -2 \,f_{T,7}
- 16\,f_{T,8} - 12\,f_{T,9})\,s^2_{2W}
- 4\,({f_{T,4}+} 2\,f_{T,6}+\,f_{T,7})\,c_{2W}^2)/64$ \\
&$++--$&
${ ((3+X^2)\,((4\,f_{T,0} + 4\,f_{T,1} + f_{T,2} +  f_{T,3}+ 2\,f_{T,4}
  - 8\,f_{T,5} -2\,f_{T,7}
  + 16\,f_{T,8} +4\,f_{T,9})\,s^2_{2W}+4(f_{T,4}+2\,f_{T,6})\,c^2_{2w})}$\\&&
${ +8(-2\,f_{T,4}+4\,f_{T,5}-2\,f_{T,6}+f_{T,7}))/32}$ \\
  &$++++$&
  $ ((-4\,f_{T,0} - 4\,f_{T,1} - 3\,f_{T,2} {-3\,f_{T,3}-2\,f_{T,4}}
  + 8\,f_{T,5} -2 \,f_{T,7}
- 16\,f_{T,8} - 12\,f_{T,9})\,s^2_{2W}
- 4\,({f_{T,4}+} 2\,f_{T,6}+\,f_{T,7})\,c_{2W}^2)/16$ \\
 \hline    
 $Z\gamma \rightarrow Z\gamma$
 &$+--+$&
 $ -X_M^2\,((-4\,f_{T,0} - 4\,f_{T,1} - 3\,f_{T,2}{- 3\,f_{T,3}+2\; f_{T,4}}
 - 8\,f_{T,5} +2 \, f_{T,7}
 + 8\,f_{T,6}  - 
 16\,f_{T,8} - 12\,f_{T,9})\,s^2_{2W}$\\&&
$ - 4\,({f_{T,4}+}4\,f_{T,5}+f_{T,7})\,c^2_{2W})/32$ \\
&$+-+-$&
 $ -X_P^2\,((-4\,f_{T,0} - 4\,f_{T,1} - 3\,f_{T,2}{- 3\,f_{T,3}+2\; f_{T,4}}
 - 8\,f_{T,5} +2 \, f_{T,7}
 + 8\,f_{T,6}  - 
 16\,f_{T,8} - 12\,f_{T,9})\,s^2_{2W}$\\&&
$ - 4\,({f_{T,4}+}4\,f_{T,5}+f_{T,7})\,c^2_{2W})/32$ \\
&$++--$&
${ ((3+X^2)\,((4\,f_{T,0} + 4\,f_{T,1} + f_{T,2} +  f_{T,3}- 2\,f_{T,4}
  - 4\,f_{T,6} -f_{T,7}
  + 16\,f_{T,8} +4\,f_{T,9})\,s^2_{2W}+(-8\,f_{T,5}+ f_{T,6}+f_{T,7})\,c^2_{2w})}$\\&&
 ${ + 4\,X_P\,(2\,f_{T,4}-4\,f_{T,5}+2\,f_{T,6}+f_{T,7}))/16}$ \\
     &$++++$&     $
 ((4\,f_{T,0} + 4\,f_{T,1} + 3\,f_{T,2} {+3\,f_{T,3}-2\,f_{T,4}}
 - 8\,f_{T,5} -2\,f_{T,7}
     + 16\,f_{T,8} + 12\,f_{T,9})\,s^2_{2W} 
+ 4\,({f_{T,4}+}2\,f_{T,6}+f_{T,7}) \, c^2_{2W})/8$ \\
\hline
    $Z\gamma \rightarrow \gamma\gamma$ 
&$+--+$&
$ {X_M^2((3\,f_{T,4}+
4\,f_{T,5} + 4\,f_{T,6} + 3\,f_{T,7})\,c_{2W}
- (16\,f_{T,8} + 12\,f_{T,9})\,c^2_W+ 
(4\,f_{T,0} + 4\,f_{T,1} + 3\,f_{T,2} + 3\,f_{T,3}
)\,
s_W^2)\,s_{2W}/(16\,\sqrt{2})}$ \\
&$+-+-$&
$ {X_P^2((3\,f_{T,4}+
4\,f_{T,5} + 4\,f_{T,6} + 3\,f_{T,7})\,c_{2W}
- (16\,f_{T,8} + 12\,f_{T,9})\,c^2_W+ 
(4\,f_{T,0} + 4\,f_{T,1} + 3\,f_{T,2} + 3\,f_{T,3}
)\,
     s_W^2)\,s_{2W}/(16\,\sqrt{2})}$ \\
     &$++--$& $ {(3 + X^2)\,((f_{T,4}+4\,f_{T,5} + 4\,f_{T,6} + f_{T,7})c_{2W} - (16\,f_{T,8} + 4\,f_{T,9})\,c_W^2 + 
    (4\,f_{T,0} + 4\,f_{T,1} + f_{T,2}+f_{T,3})\,s_W^2)\,
   s_{2W}/(8\,\sqrt{2})}$ \\
&$++++$&
$ {((3\,f_{T,4}+
4\,f_{T,5} + 4\,f_{T,6} + 3\,f_{T,7})\,c_{2W}
- (16\,f_{T,8} + 12\,f_{T,9})\,c^2_W+ 
(4\,f_{T,0} + 4\,f_{T,1} + 3\,f_{T,2} + 3\,f_{T,3}
)\,
s_W^2)\,s_{2W}
/(4\,\sqrt{2})}$ \\
\hline      
$\gamma\gamma \rightarrow \gamma\gamma$
&$+--+$&
$ -X_M^2\,(-2\,({3\,f_{T,4}+}
4\,f_{T,5} + 4\,f_{T,6} + 3\,f_{T,7})\,c_W^2 + 
2\,({3\,f_{T,4}+}4\,f_{T,5} + 4\,f_{T,6} + 3\,f_{T,7} - 8\,f_{T,8} - 6\,f_{T,9})\,c_W^4 $\\&&
$- (4\,f_{T,0} + 4\,f_{T,1} + 3\,f_{T,2}{+3\,f_{T,3}})\,s_W^4)/16$ \\
&$+-+-$&
$ -X_P^2\,(-2\,({3\,f_{T,4}+}
4\,f_{T,5} + 4\,f_{T,6} + 3\,f_{T,7})\,c_W^2 + 
2\,({3\,f_{T,4}+}4\,f_{T,5} + 4\,f_{T,6} + 3\,f_{T,7} - 8\,f_{T,8} - 6\,f_{T,9})\,c_W^4 $\\&&
$- (4\,f_{T,0} + 4\,f_{T,1} + 3\,f_{T,2}{+3\,f_{T,3}})\,s_W^4)/16$ \\
&$++--$& $-(3 + {X^2})\,(-2\,({f_{T,4}}+
4\,f_{T,5} + 4\,f_{T,6} + f_{T,7})\,c_W^2 + 
2\,({f_{T,4}+}4\,f_{T,5} + 4\,f_{T,6} + f_{T,7} - 8\,f_{T,8} - 2\,f_{T,9})\,c_W^4
$\\ &&
$- 
     (4\,f_{T,0} + 4\,f_{T,1} + f_{T,2}{+f_{T,3}})\,s_W^4)/8$ \\
     &$++++$ & {$ {((6\,f_{T,4}+8\,f_{T,5} + 8\,f_{T,6} + 6\,f_{T,7}
       )\,c_W^2 + 
       2(- 3\,f_{T,4}-4\,f_{T,5} - 4\,f_{T,6} - 3\,f_{T,7}
       + 8\,f_{T,8} + 6\,f_{T,9})\,c_W^4}$}\\ &&
     ${+ 
  (4\,f_{T,0} + 5\,f_{T,1} + 3\,f_{T,2}+3\,f_{T,3})\,s_W^4)/4}$\\
\hline
\end{tabular}
\caption{Continuation of Table ~\ref{tab:ampt1}}
\label{tab:ampt2}
  \end{table}
\endgroup

\begin{table}[h]
\begin{tabular} {|c|c|}
\hline
&$\displaystyle
\times g^4\frac{s^2}{M_W^4}$\\\hline
$W^+W^+\rightarrow W^+W^+$
& $((5 +{ X^2})\,f_{P,11} + 2\,(1 + {X^2})\,f_{P,6})/16$\\
$W^+Z\rightarrow W^+Z$
& $((5 + {2\,X+ X^2})\,(f_{P,11}+f_{P,24}) + 2\,{(1 -X)^2})\,
(f_{P,23}+f_{P,6})/16$\\
$W^+W^-\rightarrow W^+W^-$
&$((7 + {2\,X+3\,X^2})\,f_{P,11} - 2\,(-5{+2\,X - X^2})\,f_{P,6})/16$\\
$W^+W^-\rightarrow ZZ$
&${( (1+X^2)(f_{P,11} + f_{P,24}) + 4\,(f_{P,23} + f_{P,6}))/ (8\,\sqrt{2})}$\\
$ZZ\rightarrow ZZ$
&$ ((3 + {X^2})\,(f_{P,11} + 2\,f_{P,23} + 2\,f_{P,24} + 4\,f_{P,26} + f_{P,6}))/8$\\
\hline
\end{tabular}
\caption{Same as Table~\ref{tab:amps} but for the operators 
  in the nonlinear representation of the electroweak symmetry
(Eqs.~\eqref{eq:p41} and \eqref{eq:p42}).}
\end{table}

\end{document}